# Using Self-Determination Theory to Design to Support Young People's Online Help-Seeking

Claudette Pretorius*

School of Computer Science, University College Dublin, claudette.pretorius@ucd.ie

David Coyle

School of Computer Science, University College Dublin, d.coyle@ucd.ie

The application of Self-Determination Theory to understand online help-seeking and the design of online help-seeking technologies presents an interesting avenue for investigation. Improving motivation to engage in the help-seeking process could be achieved using the Basic Psychological Needs Theory as a structure to guide the design of online help-seeking technologies and online mental health resources. Positive online help-seeking experiences have an important role to play in sustained help-seeking and improved health outcomes.

CCS CONCEPTS • **Human-centered computing** •**Human-centered computing~Human computer interaction (HCI)~HCI design and evaluation methods~User studies**

**ACM Reference Format:**

Claudette Pretorius, and David Coyle. 2022. Using Self-Determination Theory to Design to Support Young People's Online Help-Seeking. In CHI '22: ACM Workshop on Self-Determination Theory in HCI: Shaping a Research Agenda, April 30– May 05, 2022, New Orleans. ACM, New York, NY, USA, NOTE: This block will be automatically generated when manuscripts are processed after acceptance.

## 1 INTRODUCTION

Mental illness has been identified as a leading cause of disability worldwide, accounting for a large portion of the burden of disease [18]. Recent studies indicate that globally, 37% of the world's young people (aged 15-24) have access to the internet [17], with searches for anxiety and depression taking place every 3-5 seconds [21].

Help-seeking has been found to be an important coping strategy and protective factor in mental illness [3]. The availability of technology and the accessibility of the internet has created the opportunity for more online support and sources of information, which offer alternatives to traditional help-seeking pathways [7]. Evidence suggests that many young people search for help online before interacting with formal mental health services, highlighting the need to understand how young people search to support their mental health and how this process can be supported [1]. The Internet provides a multitude of mental health related content, including formal sources of help and information, from verified sources such as government health websites, as well as informal sources of information and help, such as personal blogs and social media accounts. When experiencing mental health distress many people use the Internet to understand their experience [4].

---

* Place the footnote text for the author (if applicable) here.

Whilst current literature examines online mental health interventions, credibility, and reliability of online information, as well as young people's attitudes towards searching for help online, a unifying model to understand young people's online help-seeking needs is lacking. Self-Determination Theory (SDT), a theory well applied in the design of positive technologies, provides the mechanisms through which to understand how a help-seeker might be motivated to engage with the help-seeking process and how to design for positive online help-seeking experiences.

## 2 BACKGROUND

### 2.1 Help-Seeking

Help-seeking can be understood as an adaptive coping process with the goal of addressing a mental health difficulty by reaching out or seeking external assistance [15]. A positive help-seeking experience encourages future help-seeking and contributes to improved health outcomes [5,9]. Similarly, the opposite is true, negative experiences of help-seeking, personal or that of someone known to the young person, is likely to decrease the likelihood of future help-seeking [8,19]. Unfortunately, the World Health Organisation [20] has highlighted that fear of stigma and discrimination means that many people experiencing mental health difficulties are reluctant to seek help [14]. Previous research has also highlighted that most young people have a preference towards self-reliance [6,11].

### 2.2 Existing Help-Seeking Models

Several theoretical models have been applied to understand help-seeking including the Theory of Planned Behaviour, the transtheoretical model of behaviour change and Rickwood's help-seeking model. For the purposes of this paper, we will focus on Rickwood's model which was specifically developed for young people as a result of a number of investigations into young people's help-seeking behaviour

*2.2.1 Rickwood's Help-Seeking Model*

Rickwood et al. [14] identified several areas within the field of help-seeking that lacked a cohesive approach and understanding, one of which being a unifying theory of help-seeking. As opposed to previous theories, they developed a theory that looked at the micro-level factors that would either promote or hinder the help-seeking process. Rickwood et al. [14] put forward a process model which is based on the understanding that help-seeking is a deeply personal journey, moving from the intrapersonal to interpersonal in an attempt by the person to address their personal or emotional concerns.

The process is initiated when the young person becomes aware that there are symptoms or difficulties that are causing them concern; they then make the appraisal whether these concerns might require intervention. The next step requires the young person to have the skills or ability to express their difficulty in a way that is understandable to others. The third step in the process requires that there be sources of help available that suit the young person's needs and the final step requires the young person to be willing to approach these sources of help and disclose their concerns. There are factors at each of these steps that can either facilitate or hinder the process. For example, if the young person has low levels of mental health literacy, they might not be able to appraise their difficulties correctly and will not move forward with the process.

## 3 DESIGNING TO FACILITATE HELP-SEEKING USING SELF-DETERMINATION THEORY

Rickwood's help-seeking model has been used to understand the intrapersonal process that takes place when a young person seeks help. This model provides a helpful foundation from which to begin to understand help-seeking, however the online setting requires consideration of the specific mediators of the online help-seeking process.

SDT has successfully been applied to development of various technologies [10]. A macro theory of motivation and wellbeing, consisting of several sub-theories. One of these sub-theories, Basic Psychological Needs Theory, proposes that each human being has three basic psychological needs: competence, relatedness and autonomy; and when these needs are satisfied, individuals are moved towards wellbeing and sustained motivation [2]. This theory states that when these needs are met, individuals are more likely to be motivated to pursue their goals [16]. SDT has successfully been applied to development of various technologies and designing to satisfy these basic psychological needs can be linked to improved engagement and sustained motivation towards goals [10].

Motivation and more specifically internal motivation are important factors to consider when designing to promote and encourage continued help-seeking. Technologies that aim to promote and support help-seeking, need to be designed to facilitate these needs instead of frustrating them. The online mental health resources to which help-seeking technologies refer, also need to be designed to meet these needs.

Help-seeking technologies need to offer choices in terms of content, opportunities to connect and activities to relieve distress, to produce search results that are meaningful and relevant. This will meet the help-seeker's need for autonomy. Our research indicates that help-seekers experience frustration when presented with an abundance of search results that did not seem to be relevant to them [13]. Similarly, their autonomy would be frustrated if the technology did not present enough resources to choose from. Autonomy can therefore be met by providing meaningful mental health related choices. It can be further facilitated by providing opportunities for young people to choose their pathway to care and determining their own help-seeking journey.

Competence plays a key role in help-seeking as it requires a knowledge of self, an awareness of available resources and knowing how to communicate one's difficulties to sources of help, both formal and informal. A help-seeking technology that supports both exploratory and focused search will support help-seeker's need for competence. Those with lower levels of mental health literacy are likely to start with exploratory searches, if their searches are satisfactory, meet their needs and improve their mental health literacy, they are more likely to progress to more focused searches. A help-seeking technology that facilitates both these types of searches will meet help-seeker's needs for competence. Additionally, an online resource that is designed well and is easy to use, will facilitate the help-seeker's need for competence. The way mental health information is presented will also impact upon help-seeker's sense of competence. Information that is overly medicalised or presented in an academic format is unlikely to meet the average help-seeker's need for competence or positively contribute to their understanding of the difficulty they are experiencing.

Relatedness can be met in various ways in the online context. The offline context requires interpersonal exchanges in help-seeking which is not necessarily the case in online settings. A social exchange does not necessarily result in a help-seeker's need for relatedness being met, to do this the exchange would have to foster connection and make the help-seeker feels valued. Findings from previous studies have shown that the need for interpersonal interaction takes place on a continuum, with some help-seeker's requiring more interpersonal interaction than others [13]. For those requiring more interpersonal interaction, the need for relatedness in the online context can be met through online chat, discussion forums or email. For help-seekers who require less interpersonal interaction, their need for relatedness can be met through elements such as personal stories presented in videos, blogs, or interviews [12,13]. Including a variety of content with personal stories or resources that provide the opportunity to connect in different way, create greater opportunity to meet help-seeker's need for relatedness.

However, certain design choices may simultaneously meet and frustrate a need. For example, anonymity has been highlighted as an important requirement in facilitating their autonomy [13]. However, anonymity can potentially limit the degree to which results, and content suggested by a help-seeking technology is personally relevant. Certain demographic details, such as age or location, are needed from online help-seekers to provide tailored resource recommendations. When a technology is unable to provide tailored resource recommendations, help-seeker's need to for autonomy will be frustrated.

## 4  CONCLUSION

This position paper proposes that Self-Determination Theory is a useful framework through which to understand young people's online help-seeking behaviours and how technology and resources designed using SDT principles can be used to assist young people to achieve their help-seeking goals. When considering online help-seeking in terms of SDT, the online environment can provide need satisfaction with an immediacy and density that is unparallel to the offline context and therefore it is an important area of investigation.